\documentclass[journal]{vgtc}                
\ifpdf
  \pdfoutput=1\relax                   
  \usepackage{graphicx}                
  \DeclareGraphicsExtensions{.pdf,.png,.jpg,.jpeg} 
\else
  \usepackage{graphicx}                
  \DeclareGraphicsExtensions{.eps}     
\fi%

\graphicspath{{figures/}{pictures/}{images/}{./}} 

\usepackage{microtype}                 
\PassOptionsToPackage{warn}{textcomp}  
\usepackage{textcomp}                  
\usepackage{mathptmx}                  
\usepackage{times}                     
\usepackage{cite}                      
\usepackage{tabu}                      
\usepackage{booktabs}                  

\usepackage{amsmath}
\usepackage{amsfonts}
\usepackage{amssymb}
\usepackage{multirow}
\usepackage{subfig}
\usepackage{graphicx}
\usepackage{makecell}

\usepackage{xcolor}

\usepackage{balance}

\newcommand{\revision}[1]{\textcolor{black}{#1}} 

\onlineid{0}

\vgtccategory{Research}
\vgtcpapertype{please specify}

\title{MeetCues: Supporting Online Meetings Experience}

\author{Bon Adriel Aseniero, Marios Constantinides, Sagar Joglekar, Ke Zhou, Daniele Quercia}
\authorfooter{
\item
 Bon Adriel Aseniero is with the University of Calgary, Alberta, Canada
\item
 Marios Constantinides is with Nokia Bell Labs, Cambridge
\item
 Sagar Joglekar is with Nokia Bell Labs, Cambridge
\item
 Ke Zhou is with Nokia Bell Labs, Cambridge
\item
 Daniele Quercia is with Nokia Bell Labs, Cambridge
}

\shortauthortitle{Biv \MakeLowercase{\textit{et al.}}: Global Illumination for Fun and Profit}

\abstract{
The remote work ecosystem is transforming patterns of communication between teams and individuals located at distance. Particularly, the absence of certain subtle cues in current communication tools may hinder an online's meeting outcome by negatively impacting attendees' overall experience and, often, make them feeling disconnected. The problem here might be due to the fact that current tools fall short in capturing it. To partly address this, we developed an online platform---\emph{MeetCues}--- with the aim of supporting online communication during meetings. MeetCues is a companion platform for a commercial communication tool with interactive and visual UI features that support back-channels of communications. It allows attendees to be more engaged during a meeting, and reflect in real-time or post-meeting. We evaluated our platform in a diverse set of five, real-world corporate meetings, and we found that, not only people were more engaged and aware during their meetings, but they also felt more connected. These findings suggest promise in the design of new communications tools, and reinforce the role of InfoVis in augmenting and enriching online meetings. 
} 


\CCScatlist{ 
 \CCScat{K.6.1}{Management of Computing and Information Systems}%
{Project and People Management}{Life Cycle};
 \CCScat{K.7.m}{The Computing Profession}{Miscellaneous}{Ethics}
}

\vgtcinsertpkg

\begin{document}


\firstsection{Introduction}

\maketitle


Meetings are one of the most common practices in which teams and organizations can mobilize their individual members to work together in a collective effort. They bring people with different cultures and personalities together to discuss topics and share ideas, brainstorm solutions, make decisions, and resolve conflicts. \revision{While organizations devote notably large amounts of resources to facilitate and support meetings, people still feel disconnected and perceive them as unproductive~\cite{romano2001meeting}.} In the U.S. alone, Americans `enjoy' 11 million formal business meetings each day and `waste' billions in unnecessary and unproductive ones every year~\cite{attentiv}. One in every three employees considers meetings ineffective and unproductive. Ineffective meetings are bad for morale and productivity, negatively impact employees' health, and are directly linked to organizations' wasted time and resources~\cite{kauffeld2012meetings, rogelberg2012wasted}. Research in Organizational and Management science shown that not only factors such as a meeting's agenda, structure, and purpose are important to its experience, but also inclusiveness~\cite{hbr_quality_experience}, dominance~\cite{romano2001meeting}, physical comfort~\cite{hbr_air_pollution}, peripheral activities (e.g., side-talks~\cite{niemantsverdriet2017recurring}), and psychological safety of sharing and  contributing~\cite{hbr_psychological_safety} are of equal importance. 

Current communication tools promise not only to increase productivity, but also to improve overall attendees' satisfaction and experience. \revision{Such tools} facilitate communication through video, audio, and textual support~\cite{mcgregor2017more, chiu1999notelook, tucker2010catchup, barksdale2012video, esbensen2014sidebar}. However, some elements remain missing in the design space, which are linked to how people experience and perceive meetings. Hence, we set out to learn these elements, and, in turn, propose a platform that captures people's experience in meetings. We demonstrate how interactive and visual UI features facilitate \emph{`subtle back-channels'} of communication, and allow attendees to be more engaged during a meeting, reflect in real-time or post-meeting, and create awareness. In so doing, we make two main contributions:

\begin{enumerate}
\item We designed (\S\ref{sec:section4}) and developed (\S\ref{sec:section5}) a companion platform---coined \emph{`MeetCues'}---which augments a commercial communication tool with subtle back-channels of communication that reflect people's psychological experiences. 
\item We deployed (\S\ref{sec:section6}) MeetCues in a diverse set of five meetings in a corporate setting, and evaluated it qualitatively in a series of interviews. We found that our platform enabled people to feel more engaged and involved during their meetings. In the light of our findings, we discuss opportunities (\S\ref{sec:section7}) that future meetings communication tools could benefit from.
\end{enumerate}

\section{Related Work}
\label{sec:section2}

\subsection{Supporting Technologies in Meetings}
A large body of work has been focused on developing meetings' assistance through contextual information, text, audio, and video support. McGregor and Tang~\cite{mcgregor2017more} developed a speech-based agent for detecting important `items' in the spoken dialogue, Shi et al.~\cite{shi2018meetingvis} used a visual narrative-based approach for meeting summarization, \revision{and Cowell et al.~\cite{cowell2006understanding} presented ChAT to identify topics or persons of interests within multi-party conversations.} As the balance of conversational turn-taking is important for group performance~\cite{woolley2010evidence}, technologies were developed to create awareness by highlighting salient moments~\cite{bergstrom2009conversation}  \revision{and visualizing participants' contribution~\cite{bergstrom2007conversation}}, provide persuasive feedback to foster collaboration~\cite{kim2008meeting}, and assist in decision-making~\cite{ancona2019minetime}. For example, NoteLook~\cite{chiu1999notelook} exploits video streams to support note taking. Catchup~\cite{tucker2010catchup} enables late participation effectively, while Banerjee et al. provided a playback system for revisiting a recorded meeting \cite{banerjee2005necessity} . Video Threads~\cite{barksdale2012video} enables asynchronous video sharing in geographically distributed teams, and SideBar~\cite{esbensen2014sidebar} provides tools that promote social engagement through contextual support. While current tools offer an abundance of technological features, there is still a need to explore how to capture subtle communication cues that can potentially turn online meetings into a pleasant experience. Our work aims to incorporate this in the design of meetings communication tools.

\subsection{Information Visualization}
Infovis surfaces non-salient data through visual variables, and amplifies our human cognition to help us better understand data~\cite{card_readings_1999}. Thus, to support subtle communication cues, we aim to design a visualization that allows meeting attendees to \emph{engage} during a meeting, \emph{reflect} on, and be \emph{aware} of their own and their fellow attendees' experiences.  

\subsubsection{Reflection}
Baumer et al.~\cite{baumer_reviewing_2014} described \emph{reflection} as the process of reviewing previous experiences, events, and stories such that one gains insight. By providing individuals with their past data, systems help them uncover inaccurate assumptions they may have about themselves, helping them to reflect on what they should change.While many prior works look at reflection in a personal level~\cite{huang_personal_2015,li_understanding_2011}, we are also concerned with interpersonal reflection, or with how one might reflect about themselves in relation to other people. \revision{For example, Ehlen et al.~\cite{ehlen2008meeting} demonstrated interfaces for presenting automatically detected topics or action items that could be used for reflection of a past meeting.} 

\subsubsection{Awareness}
To enable people to reflect upon themselves, they must first become aware of their data~\cite{li_understanding_2011}. Previous works have focused on building techniques that create awareness and also reduce the cost of communication between collaborators. For instance, through \emph{brushing and linking}~\cite{buja1991interactive} which improves awareness by keying different areas of a visualization to each other such that a change in a particular element is reflected on all areas in which that element appears. Isenberg et al.~\cite{isenberg2010exploratory} found that collaborative brushing and linking empowered analysts to validate and build upon each other's results by avoiding redundant work. Similarly, Hajizadeh et al.~\cite{hajizadeh2013supporting} demonstrated that persistent selection created collaborators' awareness while causing minimal interference with independent work. Mahyar and Tory~\cite{mahyar2014supporting} explored the notion of Linked Common Work (LCW) to facilitate collaborative sense-making. Unlike brushing and linking, which is primarily applied to search/retrieve queries and documents, LCW incorporates users' externalizations such as recorded findings and notes. By applying the linking concept to these externalizations, they enabled analysts to understand how their findings relate to their peers, and build a common ground by helping each other to solve analytical problems. Another line of research have looked into visualizing conversations\revision{~\cite{chandrasegaran2019talktraces}} and unveiling subtle information such as emotional tone \cite{tat_visualising_2002,tat2006crystalchat}.
Vande Moere et al. \cite{moere_visualising_2008} explored team collaborations and dynamics visualizations by focusing on what ``good and poor collaboration looks like,'' which can potentially inform teams how to improve their collaborative methods. However, these previous works are made post-collaboration and are not real-time visualizations. They also rely heavily on textual information rather than the emotional exchange that occur during collaboration. Our tool builds on these previous works by visualizing each attendees' states in real-time in addition to post-collaboration support.
\section{Design Methodology and Requirements Gathering}
\label{sec:section4}

\revision{Designed iteratively, we made several versions of an interactive platform coined \textbf{MeetCues} which we tested and refined.} MeetCues (\autoref{fig:empathic_companion}) allows attendees to \emph{engage} during a meeting, \emph{reflect} on, and be \emph{aware} of their own and their peers' experiences. We illustrate our design decisions through a running example about \textit{Amalia}---a persona of a meeting attendee---and how she would use it. 

We designed five UI features (visual and interactive) to support Amalia's tasks during (\autoref{fig:empathic_companion} a--d) and after (\autoref{fig:empathic_companion} e) the meeting. Visual features portray information back to the end-users through visual UI elements (i.e., output), whereas interactive features are UI elements that allow end-users to interact with (i.e., input). Initially, Amalia joins the MeetCues Companion UI using a unique hashtag provided by the meeting host and her email address. This allows her to join a meeting room connected to an online video-conferencing service (e.g., Cisco WebEx). This is explained further in (\S\ref{sec:section5}).

\subsection{Supporting Engagement} \label{subsection:support_engagement}
\emph{Engagement} is often attributed to the extent of psychological comfort (i.e., contribution and motivation) attendees experience during their meetings~\cite{hbr_psychological_safety}. Thus, we devised a set of features that aim to make attendees feel comfortable to contribute, receive attention, and be satisfied with the usefulness of meetings. To make Amalia feel more inclined to contribute, we designed interactive features taking pointers from casual interactions \cite{mcewan_supporting_2005,pohl_focused_2013,romero_field_2007} that do not require a person's full attention \cite{pohl_focused_2013}. Since casual interactions require minimal effort, people could be inclined to do them more often. All contributions and reactions are anonymized to make attendees more comfortable in disclosing both positive and negative information \cite{joinson_self-disclosure_2001}.

\textbf{Feature 1 -- Likes \& clarify clicks (\autoref{fig:empathic_companion} c):} We developed our tagging feature by borrowing from social media cues (e.g., Facebook's like button) which has been shown to increase people's feelings of social and emotional gratification \cite{hayes_one_2016}. Amalia uses it to express her thoughts by \emph{tagging} points during the meeting with a like/clarify reaction. She does so by clicking the ``like'' button during times when she liked or agreed with what the speaker is saying, or by clicking the ``clarify'' button during times when she feels unsure about what is being said. These are casual interactions and do not require her full attention.

\textbf{Feature 2 -- Comments \& upvotes (\autoref{fig:empathic_companion} d):} To provide another channel for contribution, we implemented a comment/chat feature which is popular for online communication \cite{vronay_alternative_1999}. Amalia uses this feature to submit comments/questions without interrupting the meeting. Similar to the tagging feature, Amalia and other attendees can ``upvote'' comments, which can be sorted in chronological order or by popularity.

\subsection{Supporting Awareness and Reflection} \label{subsection:support_reflection}
\emph{Awareness} and \emph{reflection} are the processes of revealing, reviewing, and bringing together previous experiences, events, and stories, in a way that one gains insight~\cite{baumer_reviewing_2014}. In our case, we support both real-time and post-meeting reflection, and based our design on previous works in reflective informatics \cite{baumer_reflective_2015,baumer_reviewing_2014} and personal visualizations \cite{huang_personal_2015}. Visualizations allow people to explore their data, gain insights \cite{fekete_value_2008}, and reflect \cite{baumer_reflective_2015,froehlich_ubigreen:_2009,li_understanding_2011,huang_personal_2015}. We visualize attendees' reactions to what is being said during the meeting (likes/clarify clicks) in their representative emoji faces. We use techniques from personal visualizations like emojis as playful data metaphors to elicit emotional responses. Thus, MeetCues provides a means for Amalia to check her colleagues' feelings, giving her an immediate feedback on how the meeting is unfolding.

\begin{figure}
	\centering
	\includegraphics[width=\linewidth,keepaspectratio]{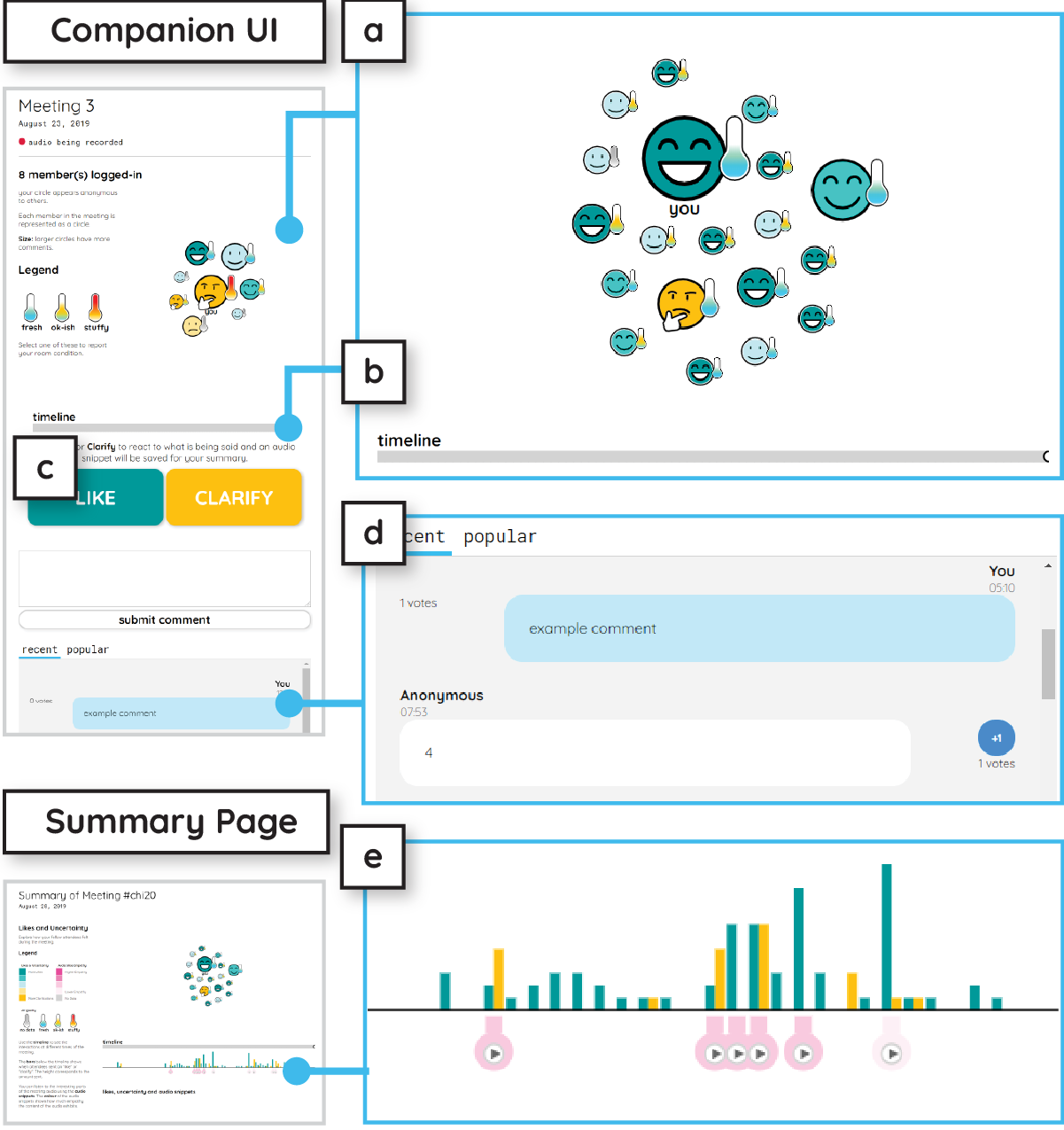}
	\caption{MeetCues' features distributed between its \textbf{companion UI} and \textbf{summary page}: (a) emoji cloud visualization - emoji faces representing attendees. The face becomes \emph{happier} and teal with more likes, and more \emph{thinking} and yellow with more clarifications, (b) timeline, (c) likes and clarify clicks, (d) comments and upvotes, and (e) audio snippets. At the end of a meeting, a summary page is sent out to attendees.}
	\vspace{-0.5cm}
	\label{fig:empathic_companion}
\end{figure}

\textbf{Feature 3 -- Emoji cloud visualization (\autoref{fig:empathic_companion} a)}: Amalia can use the \emph{emoji cloud visualization} to keep track of how the meeting is unfolding in real-time (Companion UI) or post-meeting (summary page). Each attendee is represented as an anonymous emoji face, indicating the crowdedness of the meeting. Their emoji's color changes depending on how many times they have pressed the ``like'' and ``clarify'' buttons. Amalia uses this to observe and be aware of the meeting's mood, seeing if more people are satisfied with the state of the meeting than uncertain. The size of their emoji grows when an attendee comments more. 

\textbf{Feature 4 -- Interactive timeline (\autoref{fig:empathic_companion} b):} A \emph{timeline} enables Amalia to see how the meeting's mood changes over time. When Amalia clicks on a specific point in the timeline, the emoji cloud will update to show her the state of the meeting at that time.
\textbf{Feature 5 -- Memorable moments \& Audio snippets (\autoref{fig:empathic_companion} e)}: After the meeting ends, a \emph{summary page} is automatically sent out to the attendees. In this page, Amalia can see the same emoji cloud and timeline. Below the timeline, there are bars showing the amount of reactions (likes/clarifies, comments) sent per minute. When the amount of reactions in a minute reaches a threshold, an \emph{audio snippet} is generated and displayed; we further elaborate this in (\S\ref{sec:section5}). As the name implies, audio snippets are short audio recordings taken from the whole recording of a meeting. Because these are generated based on when attendees interacted during the meeting, they can act as signs for \emph{memorable moments} of a meeting. This requires us to audio record the meeting, hence, we added a \emph{red circle} visual to alert attendees about the recording. Organizers can choose not to record their meeting which results in zero audio snippets in its summary.

\begin{figure} 
	\centering
	
	\includegraphics[width=0.95\linewidth,keepaspectratio]{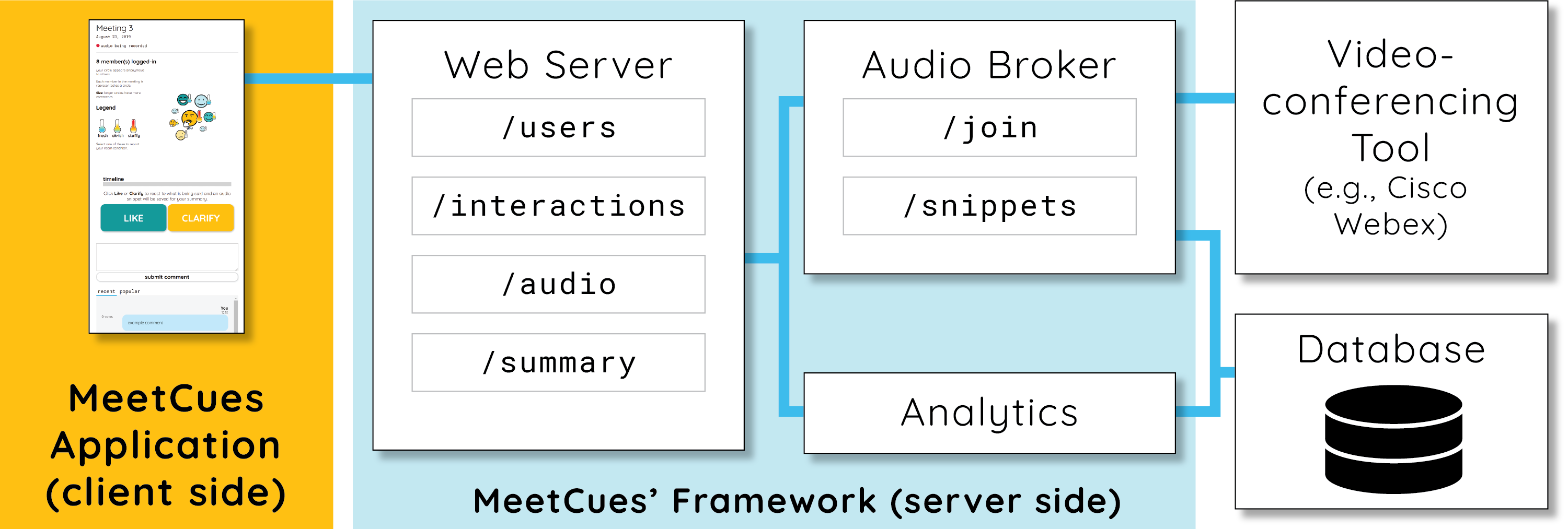}
	\caption{MeetCues system architecture.}
	\vspace{-0.5cm}
	\label{fig:architecture}
\end{figure}
\section{System Architecture and Implementation}
\label{sec:section5}

MeetCues' framework is composed by an application layer (companion UI and summary page), a web services layer (web server and audio broker), and a data access layer (\autoref{fig:architecture}). The \emph{application layer} was developed using HTML5/JavaScript and D3~\cite{bostock_d3_2011}. This allowed us to build both the visual and interactive features in a seamless way to maximize end-user's experience with the platform. \revision{The \emph{web server} was implemented using the Python Tornado framework, and exposes four endpoints for managing end-users accounts, storing and retrieving interactions with the UI, and generating summary pages.} The \emph{audio broker} was implemented using the Python Flask framework, and exposes endpoints for recording and processing meetings' audio; \revision{recording is done via Twilio's API\footnote{https://www.twilio.com/}}. The current implementation was extensively tested with Cisco WebEx\footnote{https://www.webex.com/}, but supports any remote meeting service that uses a call-in functionality to join (e.g., Zoom\footnote{https://www.zoom.us}). When a meeting ends, the audio broker stores its recordings in the file system, processes them in the background, and partitions them into \emph{audio snippets} which correspond to periods of the meeting with high engagement. It does so by computing the engagement value per 1-minute slice using the number of interactions (like and clarify clicks) and comments within the slice. It normalizes these engagement values against the whole meeting's interactions and comments, and creates the snippets for slices with values above the threshold of 0.3 (a threshold chosen after pilot deployments). It then stores these snippets in the database which can be retrieved via the web server's endpoints. We consider these snippets as `memorable moments'. Finally, the \emph{data access layer} was implemented as a MongoDB\footnote{https://www.mongodb.com/} instance, and the data is stored as database collections.

\section{Evaluation}
\label{sec:section6}
To evaluate MeetCues, we investigated its usefulness on people's experience of real-world meetings in a corporate setting. We formulated the following research question: RQ: \emph{What's the interplay between MeetCues' UI features and people's experience of meetings?}, and addressed it qualitatively by understanding the role of MeetCues in engaging people, enabling reflection, and creating awareness.

\subsection{Participants and Procedure}
We deployed MeetCues in a diverse set of five real-world meetings in a large organization, and a total of 55 users interacted with it. The five meetings included two seminar-like meetings \revision{amounting to a total duration of 190 minutes (type A) and 31 participants}. The three recurring team meetings lasted 164 minutes (type B) \revision{in which 24 participants attended}. \revision{Seminar-like meetings include information sharing in one-to-many or many-to-many fashion, while recurring ones include team status update, decision-making, or information sharing where all participants engage in the conversation.} \revision{These are two common meeting types that occur in the workplace.} To ensure diversity, we chose meetings with attendees of varying positions \revision{from students, researchers, to some with senior positions}.

We instructed our participants to log into MeetCues, which was made accessible via a public URL, using a hashtag and their email address. The first author briefly explained the aim of the platform without disclosing any detailed explanation about its features. Participants used the platform in parallel to their attendance in the meetings. 

\begin{table}
  \centering
  \begin{tabular}{lllll}
    \toprule
    \textbf{S/N} &\textbf{Age} & \textbf{Gender} & \textbf{Meeting type(s) attended} & {\textbf{\revision{Role}}}\\
    \hline
     P1 & 33 & Male & 1 type A & \revision{Researcher} \\
     P2 & 34 & Male & 2 type B and 1 type A & \revision{Senior} \\
     P3 & 26 & Male & 1 type B & \revision{Student}\\
     P4 & 34 & Male & 3 type B and 1 type A & \revision{Senior} \\
     P5 & 33 & Female & 2 type B & \revision{Researcher} \\
     P6 & 28 & Female & 1 type A and 1 type B & \revision{Student}\\
     P7 & 32 & Male & 1 type B & \revision{Student}\\
     P8 & 35 & Male & 1 type B and 1 type A & \revision{Student}\\
  \bottomrule
\end{tabular}
\caption{Demographics of interviewees.}
	\vspace{-0.5cm}
\label{tbl:interview_participants}
\end{table} 

\subsection{Qualitative Analysis and Results} \label{subsec:qual}
\revision{Following previous work recommendations on small-group user studies in real-world settings~\cite{caine2016local},} we conducted a series of interviews with eight people \revision{(who responded to an interview request)} out of the 55 who interacted with the tool (\autoref{tbl:interview_participants}).

We transcribed the interview audio, and the transcripts were then thematically analyzed~\cite{braun2006using}. A combination of open coding and axial coding was used. First, interviews were transcribed using Google's speech-to-text service, and the first author corrected the transcripts line-by-line for any disambiguation. At this stage, relevant statements were labeled. Second, axial coding was used to identify relationships between concepts and categories that emerged during open coding. Additional emphasis was given on casual interaction, visual engagement and reflection, as these were the focus of this research. \revision{Themes were reviewed in a recursive manner rather than a linear} 
by re-evaluating themes and coded text as necessary~\cite{braun2006using}. We found four high-level themes related to: \emph{Self-Expression}, \emph{Visual Engagement}, \emph{Intervention}, \emph{Sense-making and Behavioral Change}. Each theme illustrates the roles MeetCues' UI features played in our participants' meeting experiences.

\textbf{\emph{Theme 1 -- Self-Expression:}}
\revision{Tagging reactions and commenting} \textbf{(Features 1 and 2)} played a major role in engaging participants. The theme stems from how our participants used MeetCues to freely express themselves---their thoughts and reactions---during the meeting that they would possibly not have expressed otherwise. Our participants liked the anonymity and how casual the interaction is \revision{(reacting with a like or a clarify)}. P1 stated \emph{``there is a hint of informality that allows you to open up more,''} suggesting the value of casual interaction and ``cute'' visual metaphors in making people feel more comfortable to contribute more. P3 commented that \emph{``people have a lot of thoughts [sic] but don't express them explicitly so you can't see how much they agree or disagree to certain ideas, whereas [MeetCues] made me kind of aware of that.''}. Similarly, P4 stated that the tool allowed them to be \emph{``a bit more vocal about things that didn't work and why''} which he claimed to be not very adept at. Some participants also praised MeetCues' casual interactions for reacting to a meeting's content. P5 stated that meetings can feel useless \emph{``if you are a participant who is not very involved in the meeting but are just there,''} but through simple interactions, one can become more involved, even by just adding a like every now and then, feeling as if they are a part of it.

\textbf{\emph{Theme 2 -- Visual Engagement:}}
The second prominent theme is concerned with how our participants used MeetCues' visual features to assess each other's reactions, thoughts, and feelings during the meeting. \revision{The emoji cloud and timeline} \textbf{(Features 3 and 4)} enabled our participants to gain awareness of each other's state of mind through subtle cues from other members. P2 described the visual features as an \emph{``additional channel''} relating it to how seeing faces during in-person meetings can give additional information about their feelings or thoughts. When many people are involved in a meeting, it becomes hard to keep track of everyone's facial expressions. For example, P4 monitored the meeting's state through the emoji cloud visualization by watching emojis' colors and sizes change as a way to observe when people contribute and react. Similarly, P5 noted that the changing emoji faces gave her an idea of whether particular moments of the meeting were positive or not. Our participants also used MeetCues to see the reactions of others during the meeting. P2 found it useful to \emph{``know what other people think, especially what they tag.''} P3 claimed that our tool made him \emph{``feel aware''} of when people agree or disagree. This shows our participants were keen on understanding the opinions of the rest of the group during their meetings. P5's comment summarizes this utility to \emph{``get the opinion of the whole group.''} as it can be often hard to gain an idea of another person's opinions.

\textbf{\emph{Theme 3 -- Intervention:}}
The third theme is concerned with how MeetCues allowed our participants to think about changing their attitude towards others. \revision{Through the emoji cloud and timeline} \textbf{(Features 3 and 4)}, they were able to notice points where they could intervene to make the meeting's atmosphere more positive. For example, P5 said that seeing likes/clarifies through the color changes of the emoji faces, \emph{``can encourage or discourage''} attendees, especially those who are speaking. She thought of this as cues, so members can \emph{``help''} or provide encouragement when reactions' to a fellow attendee's involvement are not as positive. While most of our participants thought about intervening naturally by speaking up, it is interesting to note that P4 thought of interacting with MeetCues as a way to influence participation. He thought that whenever he hits like/clarify, \emph{``people are going to see these changes, and impact how the meeting goes.''} This is perhaps due to the fact that, while MeetCues gave our participants the means to identify when they could intervene, it is still up to them if they would do so. P4 said that he \emph{``started to pay more attention to the reactions of others''} but he did not deliberately try to change the course of the meeting, rather he simply reflected on how he should have done so more actively. 

\textbf{\emph{Theme 4 -- Sense-making and Behavioral Change:}}
The fourth theme is concerned with how our participants used MeetCues to make sense of what happened in the meeting, and to improve on or change their behavior for future meetings. \revision{The emoji cloud, timeline, and audio snippets} \textbf{(Features 3, 4, and 5)} in the summary page enabled them to reflect upon what happened during a meeting. Many of our participants (P3--P6) used terms such as ``recall'' and ``recollect'' to describe their experiences. There are three distinct sub-themes that our participants made-sense of. First, they tried to place value on the meeting by looking at the points when people reacted with a like/clarify or commented. Second, they used the audio snippets to listen to what they call as important or ``key points'' to recall what had been stated. P3 claimed that he has a \emph{``bad memory''} and used the audio snippets to recall important points. P2 and P3 defined other key points to be the memorable ones, and claimed that reviewing them could lead to more understanding of what happened during the meeting. For example, P5 said this can be used to \emph{``clarify the points... if you didn't understand something, you might do it from the comments.''} The last sub-theme is concerned with how our participants reflected on themselves to change for the better. Here, they sought areas for improvement internally (i.e., improving one's skills) and externally (i.e., improving one's self for the betterment of the group). For example, P6, who was a speaker during one recurring meeting, stated \emph{``I especially liked to replay the audio snippets... [they] can help you understand how you actually sound, what is the phrasing of your argument, and how you can improve.''} She used MeetCues to reflect on her performance and find areas of improvement. P4 stated that MeetCues can enable behavioral change in the long run by its association of key points with attendees' reactions. Through this association, meeting organizers can reorganize their groups' meetings based on people's reactions, finding points that work and improving those that are less clear.

While our participants were able to reflect on their experience, they also mentioned challenges that they encountered. Some participants found that they forgot to use MeetCues when they become immersed in the meeting (P3, P4, P5). For example, P3 said \emph{``I was listening to the conversation and I occasionally forgot to press the Like/Clarify buttons.''} Furthermore, some participants, like P4, deliberately did not use the comments section. However, while they may not contribute a comment, some still read the comments like P1 who observed that \emph{``[others] were having interesting conversations on the chat page.''}

\section{Discussion and Future Work}
\label{sec:section7}
The qualitative analysis allowed us to gain insights into the role of MeetCues' features and people's experience of meetings. The themes we uncovered illustrated how our participants used our tool, and highlighted its strengths and weaknesses. On the upside, first, by implementing casual interactions such as tagging and commenting (Features 1 and 2), our participants were able to open up and engage in expressing their thoughts during their meetings. Secondly, through the emoji cloud and timeline (Features 3 and 4), they followed what their peers are tagging and commenting on. This hints that they used MeetCues in ways that could promote empathy between meeting attendees, and actively seek to be aware of others' state of mind. \revision{Of course, future studies could further investigate the role that anonymity played in opening up.} Thirdly, those who felt the need to change a meeting's atmosphere to a more positive one suggest that they were able to become more attuned to other attendees' experiences. Finally, they were able to reflect on the meetings' contents using Features 3--5. It allowed them to make sense of their meetings and find what they consider to be important points. This suggests that our tool can potentially help attendees in contextualizing their meeting's purpose and formulate actionable points for the future. On the downside, MeetCues can be improved by providing mechanisms to help attendees get over the initial novelty of the application for continued use. \revision{For example, P3 stated that \emph{``during the first exposure to [MeetCues] I [got] curious about it and it looked interesting so maybe that could have changed...''} suggesting that they might have only used it because it was new}. MeetCues should also provide a more seamless acquisition of reactions and comments during the meeting. This is a common design problem as current tools are still not seamlessly integrated with meetings. One design idea to support this is to implement a visual cue that would remind participants of the app (e.g., by subtly dimming and brightening the application screen, or making reactions more salient but not distracting). Finally, further evaluations of MeetCues combining observations and interviews with all participants can contextualize more of our findings.

\section{Conclusion}
Communications tools promise to facilitate better communication that would yield improved perceived meetings experience; yet, people feel disconnected during online meetings. Through an iterative design process, we developed MeetCues, an online interactive platform, which allows attendees to engage during a meeting, reflect on, and be aware of their own and their peers' experience. In a qualitative study, we explored the role of our design decisions, and found that our platform enabled people to feel more involved during their meetings.



\bibliographystyle{abbrv-doi}

\bibliography{references}
\end{document}